\documentclass[12pt]{iopart}
\pdfoutput=1
\usepackage[normalem]{ulem}
\usepackage[utf8x]{inputenc}
\usepackage{ucs}
\usepackage[english]{babel}
  \expandafter\let\csname equation*\endcsname\relax
  \expandafter\let\csname endequation*\endcsname\relax
\usepackage{amsmath}
\usepackage{amsfonts,bbm}
\usepackage{amssymb}
\usepackage{graphicx,color}
\usepackage{subfigure}
\usepackage[section]{placeins}
\usepackage{hyperref}

\newcommand{\bra}[1]{\langle#1|}
\newcommand{\ket}[1]{|#1\rangle}
\newcommand{\braket}[2]{\langle#1|#2\rangle}
\newcommand{\ketbra}[2]{|#1\rangle\langle#2|}

\renewcommand{\Tr}{\mathrm Tr}

\begin{document}

\title{Entanglement critical length at the many-body localization transition} 

\author{Francesca Pietracaprina}
\address{Dipartimento di Fisica, Universit\`a degli Studi di Roma ``La Sapienza'', I-00185, Roma, Italy}
\author{Giorgio Parisi}
\address{Dipartimento di Fisica, Universit\`a degli Studi di Roma ``La Sapienza'', I-00185, Roma, Italy}
\address{INFN, Sezione di Roma I, CNR-NANOTEC UOS Roma, I-00185, Roma, Italy}
\author{Angelo Mariano}
\address{ENEA, Italian National Agency for New Technologies, Energy and Sustainable Economic Development,
viale Japigia 188, I-70126, Bari, Italy}
\author{Saverio Pascazio}
\address{Dipartimento di Fisica, Universit\`{a} di Bari, I-70126 Bari, Italy}
\address{INO-CNR, 50019 Sesto Fiorentino, Italy}
\address{INFN, Sezione di Bari, I-70126 Bari, Italy}
\author{Antonello Scardicchio}
\address{The Abdus Salam ICTP, Strada Costiera 11, I-34151 Trieste, Italy}
\address{INFN Sezione di Trieste, Via Valerio 2, I-34127 Trieste, Italy}

\begin{abstract}
We study the details of the distribution of the entanglement spectrum (eigenvalues of the reduced density matrix) of a disordered spin chain exhibiting a many-body localization (MBL) transition. In the thermalizing region we identify the evolution under increasing system size of the eigenvalue distribution function, whose thermodynamic limit is close to (but possibly different from) the Marchenko-Pastur distribution. From the analysis we extract a correlation length $L_s(h)$ determining the minimum system size to enter the asymptotic region. We find that $L_s(h)$ diverges at the MBL transition. We discuss the nature of the subleading corrections to the entanglement spectrum distribution and to the entanglement entropy.
\end{abstract}

\maketitle

\makeatletter

\section{Introduction.}
The study of disordered, isolated quantum systems has recently received a lot of attention due to both technological progress in the isolation and manipulation of mesoscopic quantum systems \cite{bloch2008many,cirac1995quantum,barends2016digitized} and theoretical advances. Among the latter, the phenomenon of Many-Body Localization (MBL) has taken center stage since its theorization \cite{basko2006metal} as the surprising breakdown of ergodicity due to quantum effects in a generic (i.e.\ non-integrable) disordered system \cite{oganesyan2007localization,nandkishore2014many}. The MBL phase, which is the continuation of the celebrated Anderson localized phase \cite{abrahams201050} to interacting systems, has been characterized, among others, via the absence of transport of conserved quantities (particles or spin and energy) \cite{basko2006metal,znidaric2016diffusive,lerose2015coexistence,nandkishore2014many}, nature of the eigenstates \cite{pal2010many,de2013ergodicity,luitz2015many}, spreading or evolution of entanglement \cite{vznidarivc2008many,bardarson2012unbounded,serbyn2013universal,goold2015total,iemini2016signatures} and emergent local integrability \cite{huse2014phenomenology,serbyn2013local,chandran2015constructing,ros2015integrals,imbrie2016review}. MBL is emerging as a compelling and universal phenomenon and its implications for our understanding of quantum dynamics and applications to future technologies are becoming evident \cite{altshuler2010anderson,knysh2010relevance,laumann2015quantum}.

Entanglement is commonly used as a characterization of a quantum state and of its non-classical properties \cite{amico2008entanglement,horodecki2009quantum}. As said before, it has been used to characterize the MBL phase as well \cite{vznidarivc2008many,bardarson2012unbounded,serbyn2013universal,goold2015total,iemini2016signatures,de2016quantum}. In disordered systems one should discuss the probability distribution over realizations of the quenched disorder of the various entanglement measures. In particular, the typical value of the entanglement entropy of a single eigenstate has been shown to be a good discriminator of the two phases: the system's entanglement follows a volume law in the ergodic phase (in line with the eigenstate thermalization hypothesis, see  \cite{srednicki1994chaos,deutsch1991quantum} and the recent review \cite{d2015quantum}) and an area law in the MBL phase \cite{pal2010many}. The sample-to-sample fluctuations of the entanglement entropy have been instead shown \cite{kjall2014many} to be good locators of the transition, and have been proposed as a signal of its first-order nature. The entanglement spectrum \cite{li2008entanglement} has proved in various situations to contain a lot more information than the entanglement entropy alone, and, where it is the case, it shows signatures of topological order. 

In the context of MBL, the entanglement spectrum has been studied in some recent papers \cite{yang2015two,PhysRevB.93.174202,serbyn2016universal} (but see also in a similar setting \cite{PhysRevB.83.045110}); when focusing on the MBL phase, these works find power-law distributions for the entanglement spectrum (similar to those found in \cite{de2013ergodicity}), while in the ergodic phase the distribution is found similar to a Marchenko-Pastur distribution, which is the distribution of eigenvalues of the reduced density matrices of random states \cite{facchi2008phase,1751-8121-40-45-017,de2010phase}; Ref.\ \cite{PhysRevB.93.174202} instead focuses on the level statistics of the entanglement spectrum, revealing interesting details of the transition on the MBL side.

The aim of this paper is to show that deviations from Marchenko-Pastur 
of the probability distribution of the entanglement spectrum in the ergodic phase provide an important characterization of the ergodic phase of such a disordered system. Moreover, such deviations from Marchenko-Pastur can be used to define a correlation length and predict the location and finite-size scaling exponents of the MBL transition. 
We will show that the entanglement spectrum therefore appears to be a crucial quantity, being able to identify even the most subtle correlations that are present in the ergodic phase of a disordered quantum system. The tools developed in this work should be useful for future studies of disordered quantum systems in their delocalized phases.

\section{Model and definitions. }
We consider what has become the \emph{standard model} of MBL in one dimension:  the spin-1/2 Heisenberg chain with random fields
\begin{equation}
 H=-J\sum_{i=1}^L \ \vec{s}_i\cdot \vec{s}_{i+1} - \sum_{i=1}^L h_i s^z_i ,
\end{equation}
where we fix $J=1$ and the random fields $h_i$ are drawn from a uniform distribution in $[-h,h]$, and $s^{xyz}=\frac{1}{2}\sigma^{xyz}$ ($\sigma^k$ being the Pauli matrices). Furthermore, we choose to work in the subspace with total magnetization $S^z=0$, whose dimension is $\mathcal N=\binom{L}{L/2}$. In the numerics we study $L=8$ to 18, and consider periodic boundary conditions. The localization transition happens, for these parameters, at $h=h_c\approx3.7$ according to various numerical investigations~\cite{pal2010many,luitz2015many,pietracaprina2016forward}.

We consider a single eigenstate of $H$, $\ket{\psi}$, computed through the exact diagonalization of $H$, so that the corresponding eigenvalue is in the center of the energy spectrum; this can be obtained by targeting an energy $E=\mathcal N^{-1} \mathrm{Tr} H$ in a shift-invert diagonalization algorithm (this estimates the mean spectral energy and accounts for its finite size corrections). Not averaging over a number of eigenstates corresponding to the same disorder realization removes a source of correlation.

To study the entanglement properties we focus on half of the chain, of size $L/2$, indicated as region $A$, the complementary region being $B$. We compute the reduced density matrix of this (pure) state as
\begin{equation}
 \rho_A=\mathrm{Tr}_B\rho=\mathrm{Tr}_B\ketbra{\psi}{\psi}
 \label{eq:rdm}
\end{equation}
and determine its $2^{L/2}$ eigenvalues $\lambda_i$; this set (or better, the set of the logarithms $\ln\lambda_i$) is called the \emph{entanglement spectrum}. Because of the $S_z=0$ constraint, (\ref{eq:rdm}) is in a block-diagonal form. We obtain numerical data with $10^6$ disorder realizations up to $L=14$, $7.5\cdot 10^4$ realizations for $L=16$ and $2,500$ to $5,000$ realizations for $L=18$.

In the delocalized phase the eigenvalues are all of the same order. Because of the normalization condition $\sum_i\lambda_i=1,$ we have therefore $\lambda_i\sim 2^{-L/2}$. One can think of the eigenvalues of order $2^{-\ell}$ as originating from states that are (approximately) uniformly entangled over a region of $\ell$ spins, so that $\log_2\lambda_i$ can be thought of as a length scale. 

A few examples will clarify the main idea. Consider first a maximally entangled state, such that the reduced density matrix of half system $A$ has all eigenvalues $\lambda_i = 2^{-L/2}$. The only scale in the system is clearly $L/2$, as entanglement is evenly spread all over subsystem $A$. As a second example, consider a state such that the maximum eigenvalue is $\lambda_1=\mu = O(1)$ and all the others are $\lambda_i= O( 2^{-L/2})$ such that $\mu + \sum_{j\neq 1} \lambda_j = 1$; then there are two scales, one $O(\log_2 \mu)=O(1)$ and the other one ``global"  $O(L/2)$. This situation typically arises when one biases the purity of a random state, so that one eigenvalue ``evaporates" out of the continuum of the other ones $O( 2^{-L/2})$ \cite{de2010phase}, carrying most entanglement. As a third example, consider a situation in which the eigenvalues split is two classes: 
$\lambda_i= O(1/M_1)=O( 2^{-L/2})$ and $\mu_j= O(1/M_2)=O( 2^{-\sqrt{L}})$, their multiplicity such that $\sum_j \mu_j + \sum_{i} \lambda_i = 1$. Then there are two scales, one more ``local", $\ell_2 = O(\log_2 M_2)=O(\sqrt{L})$ and the other one ``global"  $\ell_2 = O(\log_2 M_1)=O(L)$; these two scales will be in general intertwined in the system, in the sense that there will presumably be correlated ``islands" of size $O(\sqrt{L})$ interspersed in the system size $O(L)$.
The above examples clarify that the scales are not necessarily related to the \emph{physical} qubits, but rather to given directions in the Hilbert spaces of system $A$ (and $B$), given by the Schmidt decomposition.

The above intuition is centerpiece in the numerical methods which are based on the truncation of the reduced density matrix \cite{schollwock2005density} (used in the context of MBL in \cite{2015arXiv150901244Y}) and will be useful in the following. In the next section we will investigate the details of the distributions and, specifically, the effects that arise from increasing the system size $L$ and approaching the critical value of the disorder $h_c$. As noticed in previous work \cite{serbyn2016universal}, in the MBL phase ($h>h_c$) the existence of a many-body localization length $L_l(h)$ implies a power-law distribution for the $\lambda_i$. In this paper we show the signs of another correlation length $L_s(h)$ in the entanglement spectrum, that will appear in the sub-leading corrections for $h<h_c$, and which diverges at the critical point. The different lengths (to be) introduced in this Article are summarized in Table \ref{tab:table1}.

\begin{table}[ht!]
  \centering
  \caption{Lengths and scales.}
  \label{tab:table1}
  \begin{tabular}{| c | c | c | c | c |}
\hline
    System size & MBL length & Correlation length & boundary length scale & bulk length scale \\
    \hline
    $L$ & $L_\ell$ & $L_s$ & $\ell_1$ & $\ell_2$ \\
    \hline
  \end{tabular}
\end{table}

\section{Entanglement spectrum and its probability distribution.}
\label{sec:pdf}

In order to identify the different length scales in the $P(\lambda)$ it is useful to consider the entanglement entropy, defined as
\begin{equation}
S_A=-\Tr{\rho_A\ln\rho_A}=-\sum_i\lambda_i\ln\lambda_i .
\end{equation}
In recent years a lot of progress has been done on understanding the properties of the entanglement entropy close to a quantum phase transition \cite{calabrese2004entanglement}. In our case, it will not be possible to get too close to the MBL phase transition due to prohibitive numerical difficulties, but we think it is useful to discuss the general setup in which these kind of questions are considered.

Using the replica trick \cite{holzhey1994geometric} $S_A$ can also be written as (we shall drop henceforth the subscript $_A$ on entropies)
\begin{equation}
S=-\lim_{n\to 1}\frac{\partial \Tr(\rho_A^n)}{\partial n} \equiv -\lim_{n\to 1}\frac{\partial S^{(n)}}{\partial n}.
\end{equation}
Usually, one takes for $\ket{\psi}=\ket{E_0}$ (assuming $E_0=0$ for simplicity) the ground state of some local Hamiltonian which can be obtained as
\begin{equation}
\label{eq:gsproj}
\ket{E_0}\bra{E_0}=\lim_{\beta\to\infty} e^{-\beta H}.
\end{equation}
Physically, one needs $\beta\gtrsim 1/\Delta_0$, where $\Delta_0$ is the gap on the ground state. One can say that $1/\Delta_0$ defines a characteristic (imaginary) time after which the properties of the ground state can be extracted from the (imaginary) time dynamics.

Since
\begin{equation}
\rho_A(s'_A|s_A)=\sum_{s_B}\braket{{s'}_A ; s_B}{E_0}\braket{E_0}{s_A ; s_B},\quad\ket{s_A ; s_B}\equiv\ket{s_1,\dots,s_{L_A};s_{L_A+1},...,s_L},
\end{equation}
we can consider $\rho_A(s'_A|s_A)$ as the transition amplitude from configuration $\{s_A\}$ to configuration $\{s'_A\}$, for the imaginary time evolution $e^{-\beta H}$.  Following this construction, one can think of $S^{(n)}=\Tr(\rho_A^n)$ as a sum over all possible configurations of spins $s_{i}$ arranged on a $n$-sheeted (lattice) surface with two conical singularities $\pm 2\pi n$ at the boundaries between $A$ and $B$. At criticality when one of the parameters $g\to g_c$, a correlation length $\xi$ diverges, and one sees scale invariance in the limit $1\ll L\lesssim \xi$. For a conformal field theory (CFT) $\xi\sim 1/\Delta\sim |g-g_c|^{-\nu}$ where $\nu$ is some critical exponent characteristic of the theory. 

Whether the continuum limit yields a CFT or not, we can define \emph{twist operators} \cite{holzhey1994geometric,calabrese2004entanglement}, which create conical singularities at $0$ and $L_A$ in the continuum limit, and write the entanglement entropy as a correlation function of these operators:
\begin{equation}
S^{(n)}(L_A)= \Tr(\rho_A^n)=\langle O^n(0)O^{-n}(L_A)\rangle ,
\end{equation}
and by taking derivatives wrt to $n$ one obtains the entanglement entropy. Notice that by knowing $S^{(n)}$ for every $n$ one can reconstruct the entanglement spectrum (and \emph{vice versa}).

If the energy $E$ of the eigenstate is extensively away from the ground state, we need to modify (\ref{eq:gsproj}) as 
\begin{equation}
\ket{E}\bra{E}=\lim_{T\to\infty}\int_{-T}^T \frac{dt}{2T}e^{-it(H-E)} + O\left( \frac{1}{T} \right),
\end{equation}
where now $T\gtrsim1/\Delta$, the gap around the energy $E$, as one can easily show by looking at the matrix elements $\braket{s'}{E}\braket{E}{s}$ \footnote{Other choices are possible, such as for instance 
$\propto e^{-\beta(H-E)^2}$ for $\beta\to\infty$. All options should incorporate some kind of evolution for long times, in order to isolate a single eigenstate.}.  
Now however, in a generic many-body model $\Delta\sim e^{-L \sigma(E)}$ is exponentially small in $L$ ($\sigma(E)$ is the entropy per spin at energy $E$) so we see that the scaling region requires $\xi\sim \ln(1/\Delta)\sim |g-g_c|^{-\nu}$, which is probably the origin of claims of a dynamical exponents $z=\infty$ for the MBL transition \cite{pal2010many}.

\begin{figure}[tbp]
\includegraphics[width=.8\linewidth]{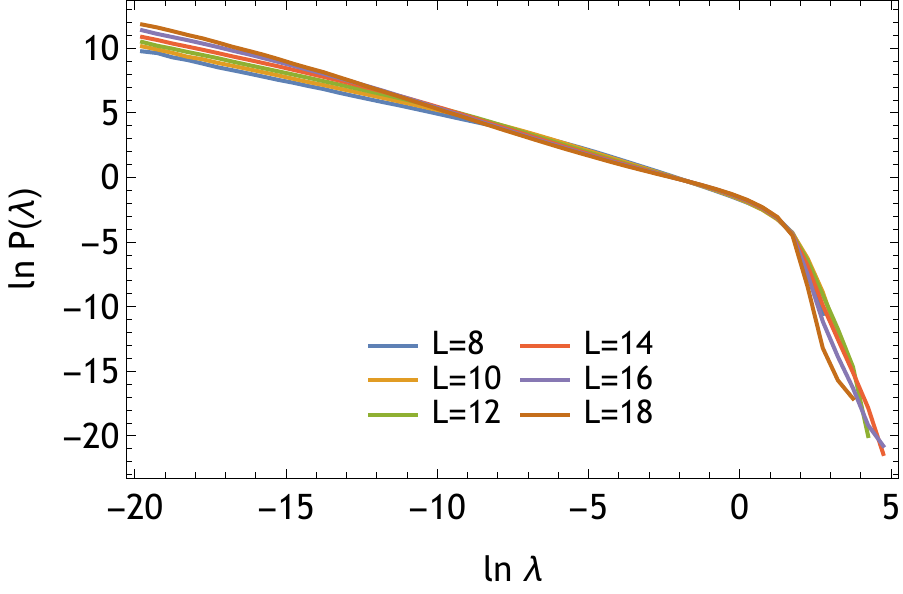}
\centering
\caption{(Color online) Rescaled entanglement spectrum $\ln P (\lambda)$ for different system size; $h=1$.}
\label{fig:P}
\end{figure}

As one takes larger and larger subsystems $L_A$ we have a change of behavior when $L_A\simeq \xi$, entering the scaling region. We now go back to the description in terms of the eigenvalues of $\rho_A$ and in the rest of the paper we will see how to identify the lengthscale $\xi$ in the entanglement spectrum. 
As said before, in the delocalized phase the eigenvalue probability distribution will have a finite support and a mean value that scales as $O(2^{-L_A})$. We can then use the system-size rescaled eigenvalues $2^{L_A}\lambda_i\to \lambda_i$, so that the mean $\langle{\lambda}\rangle=1$ does not scale with the system size. Typically, one observes (scaled) distributions such as those displayed in Fig.\ \ref{fig:P} (obtained for $h=1$).
If this scaling is observed, then it is easy to show that 
\begin{equation}
\Tr(\rho_A^n)=S^{(n)}(L_A)=2^{-(n-1) L_A}\phi_n(L_A).
\end{equation}
By taking derivatives, the entanglement entropy reads
\begin{equation}
\label{eq:SL}
S=\ln 2\ L_A-s_{L_A},
\end{equation}
where $L_A$ is the length of the interval $A$. The second term is related to the function $\phi_n$ and contains non-trivial information on the correlations between the twist operators. In particular it may also contain a non-trivial length scale. Notice that $s_{L_A}\geq 0$ since the first term on the right-hand side is also the maximum possible value of $S_A$. As we have chosen $L_A=L/2$, we will denote all the quantities with their $L$ (rather than $L_A$) dependence. We will also indicate with $\lambda$ the rescaled eigenvalues of the reduced density matrix.

\section{Deviations from Marchenko-Pastur and emergence of other scales.}
\label{sec:deviations}

Previous works~\cite{yang2015two,serbyn2016universal} noticed that in the delocalized phase, a substantial part of entanglement spectrum is well approximated at low disorder by a Marchenko-Pastur (MP) distribution
\begin{equation}
 P(\lambda)=\frac{1}{2 \pi} \sqrt{\frac{4 - \lambda}{\lambda}}.
 \label{eq:mp}
\end{equation}
For MP, the constant in Eq.\ (\ref{eq:SL}), in the thermodynamic limit, is $s_\infty=1/2$. However, finite-$L$ corrections are important, and, if we assume that the corrections follow the same pattern as those in the MP distribution we should find  
\begin{equation}
\label{eq:sLMP}
s_L\simeq s_\infty+c\ e^{-m L/2},
\end{equation}
where for MP $s_\infty=1/2$, $m=\ln 2$, $c\simeq 0.38(3)$ ($m$ and $c$ come from the exact diagonalization of random Wishart matrices: an explicit calculation should be possible using methods of random matrix theory \cite{facchi2008phase,majumdar2009index,de2010phase} and is left for future work). Notice that Eq.\ (\ref{eq:sLMP}) is of the form of the corrections expected from quantum field theory (see \cite{doyon2009bipartite} and references therein) where $m$ is the mass of the lightest excitations. Using this form for fitting the numerical data we find $s_\infty\simeq 0.66(1)$ and $m\simeq 0.20(1)$ (around $h=1$, with a small dependence on $h$), therefore making evident that MP is \emph{not} the limit of the rescaled eigenvalues distribution. In order to collect more evidence about such difference and study the evolution of the entanglement spectrum at the MBL (critical) transition point, we need to look at the whole distribution.

Although Eq.\ (\ref{eq:sLMP}) might suggest a fast approach to the thermodynamic limit, we remark that the asymptotic limit is achieved only when $L$ is larger than the correlation length $\xi$ discussed before, which diverges at the transition at $h=h_c=3.7$. We will see how to identify this length, which we will call $L_s(h)$, in deference to the appearance of a \emph{spinodal point} in the entanglement spectrum,
and show that $L_s\gtrsim 20$ already at $h=2.5,$ representing therefore the main roadblock to the observation of a true asymptotic region in current numerics. The width of the numerically accessible critical length demands that we understand more accurately the sub-leading corrections to the entanglement quantities in MBL, if we want some more information on the theory of the critical point.

The remainder of this paper is focused on the definition and study of this correlation length $L_s(h)$ in the delocalized region, by scrutinizing its divergence and the associated finite-size corrections.
 
 \begin{figure}[tbp]
 \includegraphics[width=.8\linewidth]{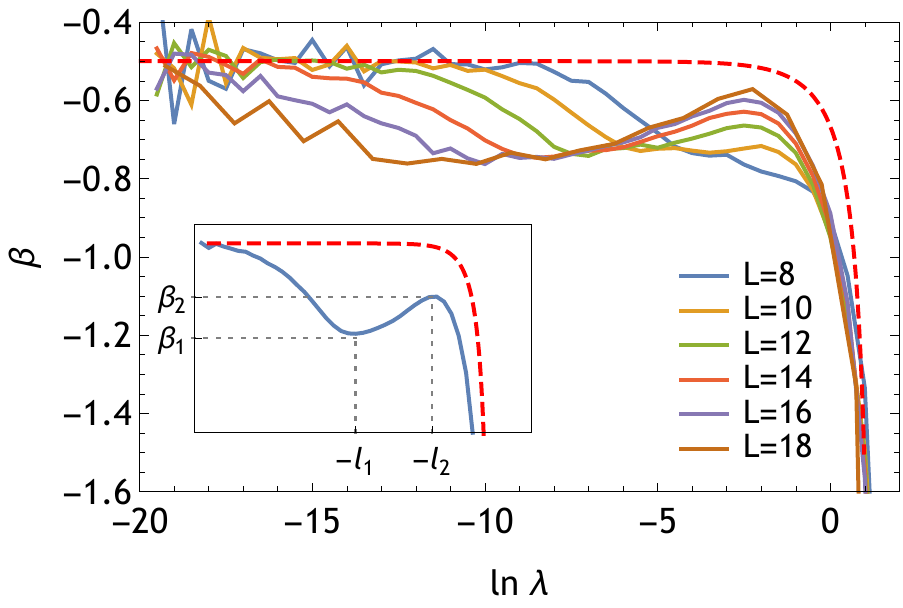}
 \centering
 \caption{(Color online) Logarithmic derivative $\beta$ at $h=1$. The logarithimc derivative of the MP distribution---dashed (red) line---is shown for comparison. 
Notice the presence of a minimum and a maximum for $L \geq 12$. We extract their coordinates by polynomially interpolating the numerical data, as exemplified in the inset. The asymptotic region of approach to the thermodynamic limit is obtained after the birth of a length scale $L_s$ where a tricritical point is observed, which in this figure is $L\simeq 10$. Similar (although less clean) results can be obtained by turning on a finite transverse magnetic field which breaks the conservation of the total magnetization $S_z$.}
 \label{fig:beta}
 \end{figure}

To better characterize the distribution we consider its logarithmic derivative
\begin{equation}
\beta(\ln\lambda,h,L)=\frac{d\ln P}{d\ln \lambda},
\end{equation}
as a function of $\ln\lambda.$
For the MP distribution one would have $\beta=-2/(4-\lambda)$ and, in the limits  $\log\lambda\to-\infty,\,\ln 4$, one would get $\beta\to -1/2,\,-\infty$, respectively. However, one numerically observes the peculiar behavior in Fig.\ \ref{fig:beta} (at $h=1$) that, we will argue, extends all the way to the localization transition point $h_c=3.7$. For fixed $h$ and small $L$, $\beta$ is monotonically decreasing from $-1/2$ (at small $\lambda$) to $-\infty$ (at large $\lambda$): see for example the $L=8$ curve in Fig.\ \ref{fig:beta}. As $L$ increases, there appears a value $L=L_s(h)$ (approximately $L=10$ in Fig.\ \ref{fig:beta}) where a tricritical point is found and where a minimum and a maximum are born in $\beta$; for $L>L_s(h)$ these extremal points are located at $\ln\lambda\equiv-\ell_{1,2}$, and we denote their values with $\beta(-\ell_{1,2},L,h)\equiv\beta_{1,2}$, respectively (see inset in Fig.\ \ref{fig:beta}). 

\begin{figure}[tbp]
 \includegraphics[width=.8\linewidth]{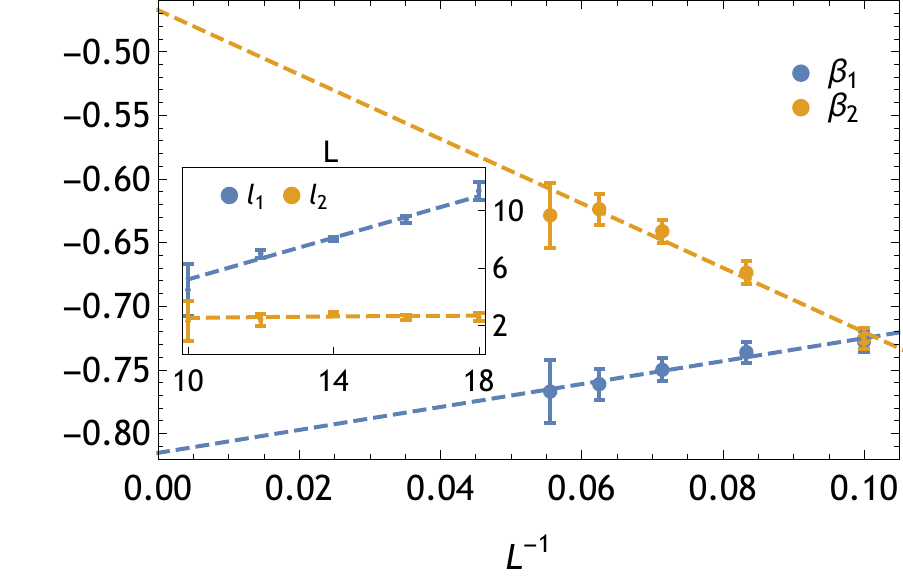}
 \centering
 \caption{(Color online) Values of the local minimum and maximum $\beta_{1,2}$, at $h=1$, as a function of the system size $L$. The value $\beta_2=-1/2$ corresponds to the MP distribution. The corresponding points of local minimum $\ell_{1}$ and maximum $\ell_{2}$ are shown in the inset.}
 \label{fig:beta12L}
 \end{figure}

The tricritical point $L_s$ is the solution of the equations (the primes denote derivatives wrt $\ln\lambda$):
\begin{eqnarray}
\beta'(\ln\lambda,L,h)=0,\nonumber\\
\beta''(\ln\lambda,L,h)=0,
\end{eqnarray}
yielding the solution $(L_s, \ln\lambda=-\ell_{1}=-\ell_{2})$ as a function of $h$. These equations can be solved easily, once a polynomial interpolating function for $\beta$ is obtained from the data. The main lesson to be learned from this observation is the existence of the length scales $L_s, \ell_1, \ell_2$. Before going to the physical interpretation of these scales we look at the evolution at fixed $h$ of the pairs $(\ell_{1,2},\beta_{1,2})$ when $L\to\infty$. As can be seen from Fig.\ \ref{fig:beta12L}, the maximum $\beta_2$ approaches the MP value, $\beta_2\to -1/2$, with its position $\ell_2$ remaining of $O(1)$ (inset); at the same time, the minimum $\beta_1$ goes to a value which is dependent on $h$ (for $h=1$, $\beta_1\to -0.8$), but its position $\ell_1$ escapes to the arbitrarily small eigenvalue region, $\ell_1\to\infty$ (inset). We offer no explanation for this value of $\beta_1$ but we notice that in our numerics, as $h\to h_c$, it seems that $\beta_1\to-1$ (similarly to \cite{de2013ergodicity}).

The distribution $P$ therefore gets closer to MP for $-\ln\lambda\geq \ell_2$ (Fig.\ \ref{fig:beta}). 
Since, as explained before, $-\ln \lambda$ is to be considered a length scale, we propose that $\ell_2$  represents a typical length scale of the system, that is related to local phenomena, and characterizes the \emph{bulk} properties of the system, describing for example the dynamics of local excitations of the eigenvector, enabling us to distinguish it from the eigenvector of a random matrix. We also surmise that the second, larger length scale $\ell_1$, which goes to $\infty$ as $L\to\infty$, defines the effects on the distribution due to the size of the cut $L_A=L/2$, hence \emph{boundary} effects. The length scale $L_s(h)$ is the minimum system size at which these two properties can be separated, and therefore it can be identified as the distance after which disturbances decay, or the \emph{correlation length} $\xi$ of our system. 

\begin{figure}[tbp]
 \includegraphics[width=.8\linewidth]{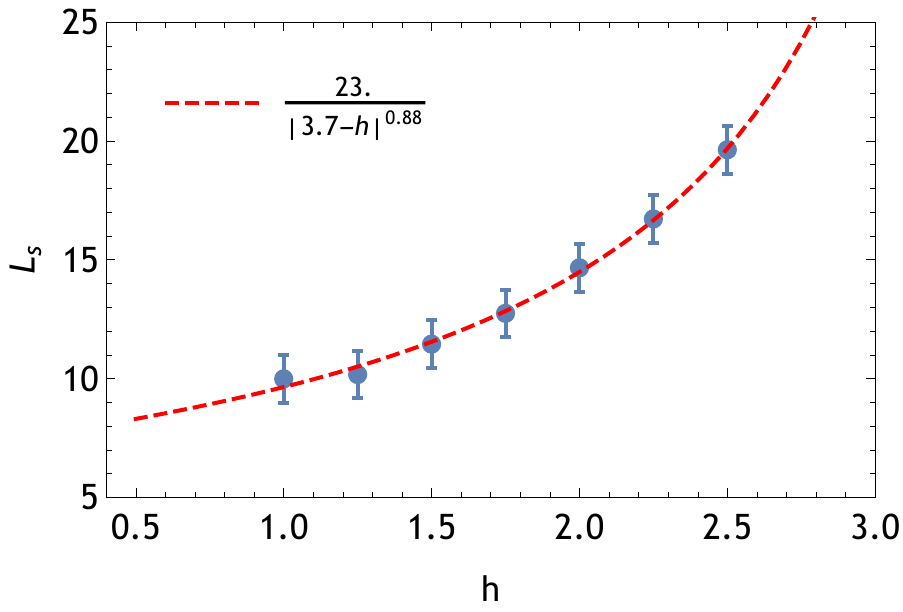}
 \centering
 \caption{(Color online) System size $L_s(h)$ at which the tricritical point is observed, as a function of $h$. The dashed line is a fit using Eq.\ \eqref{eq:Lshfit}; all the points $h\geq 1$ are used in the fit.}
 \label{fig:Lsh}
 \end{figure}

We plot $L_s$ versus $h$  in Fig.\ \ref{fig:Lsh} and observe its divergence at the MBL phase transition. The fit (dashed red line) gives a result consistent with the data in \cite{luitz2015many}:
\begin{equation}
\label{eq:Lshfit}
L_s(h)\simeq\frac{a}{(h_c-h)^{\nu}},
\end{equation}
where $h_c=3.7\pm 0.4$ and $\nu=0.9\pm 0.2$. On the other hand, by fixing the transition point at $h_c=3.72$ in the fitting procedure we get $\nu=0.88$ with smaller errors (in both cases $a\approx 23$); see Figure \ref{fig:Lsh}. Notice that $L_s$ is a proper length, measured in lattice sites. Notice also that the critical exponents are close to $1$. This value is again in agreement with what is found in the numerical literature, but inconsistent with a Harris-like bound in \cite{chandran2015finite}, which suggest $\nu\geq 2$, and renormalization group analyses \cite{vosk2015theory,potter2015universal} which give $\nu\simeq 3$. We must remark however that, upon identification of the length $L_s(h)$ which defines the beginning of the asymptotic region, the true critical exponents should be extracted by analyzing data at $L\gg L_s(h)$ and $h\lesssim h_c$. Current technology allows $L\lesssim 24$, which means $h<2.8$, therefore subleading corrections to the critical scaling might be substantial (an analog situation, in a similar context is observed in \cite{obuse2012finite}). In particular a fit of the form
$L_s\left({h}/{h_c}\right)\simeq a\,{\left(1-\frac{h}{h_c}\right)^{-3}\left[1-b_1\,\left(1-\frac{h}{h_c}\right)+b_2\,\left(1-\frac{h}{h_c}\right)^2\right]}$ is a decent alternative to (\ref{eq:Lshfit}) even with $h_c=3.72$ fixed, yielding $b_1\sim 7$ (admittedly not a small correction). This situation is unfortunate but cannot be mended in absence of either a full theory of the critical point of MBL (providing with the nature of the subleading corrections due to irrelevant operators, in the language of renormalization group analysis) or some radically new numerical tools (e.g.\ a \emph{quantum computer}), enabling the study of eigenstates/eigenvalues for significantly larger system sizes (say $L\sim 100$).

We now return to the subleading terms $s_L$ in the entanglement entropy, discussing the finite size effects in this quantity as well. For $h\lesssim h_c$, if $L\ll L_s(h)$ the system should show critical behavior. All indicators from previous works (like absence of level repulsion, statistics of eigenvalues, entanglement entropy etc.\ \cite{luitz2015many,potter2015universal,2016arXiv160601260Y,khemani2016critical}) show that the critical behavior is very similar to the MBL phase and that a jump of the entanglement entropy occurs at the transition. An analogous result is obtained also in studies of the (single particle) Anderson model on the Bethe lattice, which shows a critical behavior indistinguishable from that of the localized region \cite{mirlin1991localization,de2014anderson}. So if $L\ll L_s(h)$ the system shows localization properties. Eigenstates in the MBL phase obey an area law \cite{vznidarivc2008many,bardarson2012unbounded,serbyn2013universal}; thus, since in the localized region $S=o(L)$, in Eq.\ \eqref{eq:SL} $s_L=\frac{\ln 2}{2}L+o(L)$ and so its derivative is $\frac{\partial s_L}{\partial L}=\frac{\ln2}{2}>0$.
On the other hand, upon increasing $L$, if the finite-size corrections in $s_L$ follow the pattern of the corrections in Eq.\ \eqref{eq:sLMP} we should observe, for $L\gg L_s(h)$, a negative derivative, $\frac{\partial s_L}{\partial L}=-c\,\frac{m}{2}\,e^{-m L/2}<0$; notice that this behavior and the sign of the derivative is independent of the precise value of the coefficients. 
\begin{figure}[tbp]
 \includegraphics[width=.8\linewidth]{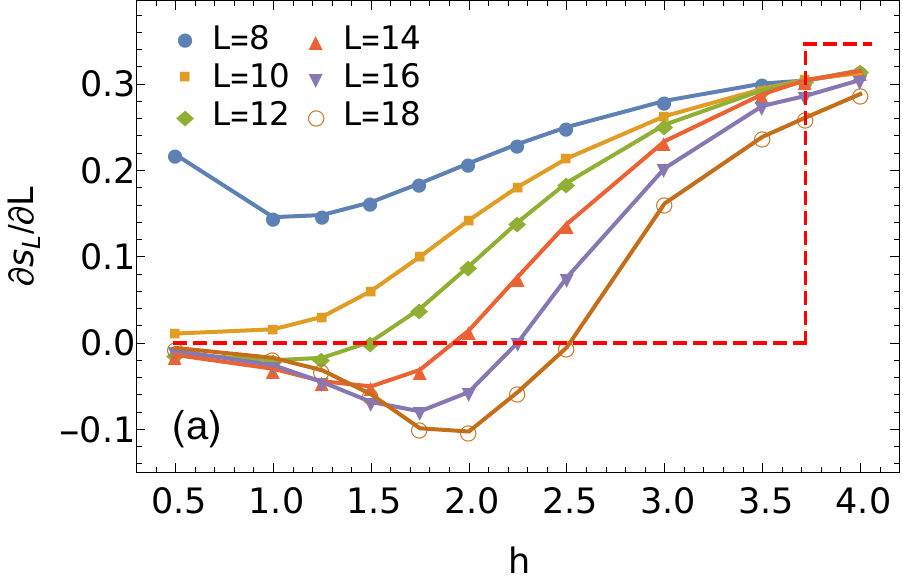}
 \includegraphics[width=.8\linewidth]{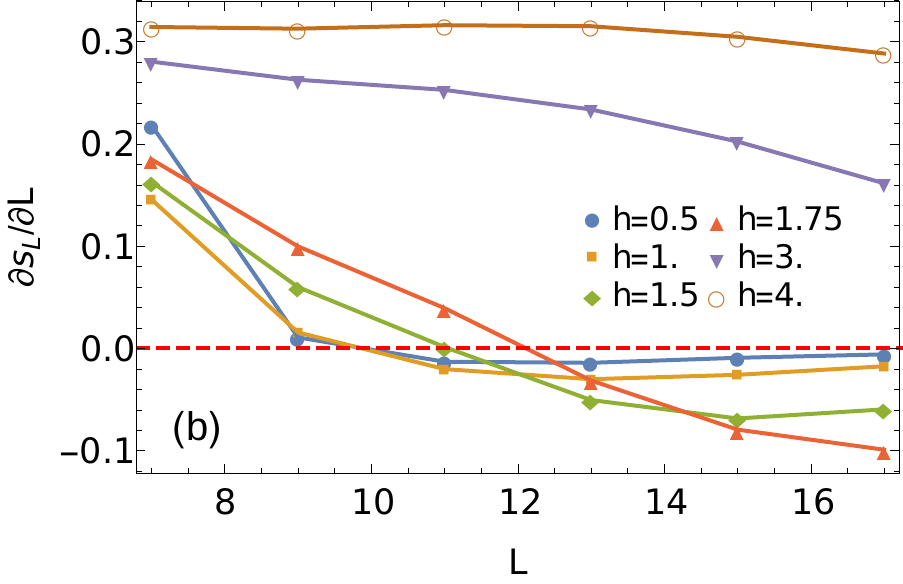}
 \centering
 \caption{(Color online) $\partial s_L(L,h)/\partial L$ calculated between $L$ and $L-2$, as a function of the disorder (a) and system size (b). The dashed red line in (a) is the thermodynamic limit, assuming that the corrections in the delocalized region are given by (\ref{eq:sLMP}).}
 \label{fig:dsdL}
\end{figure}
Summarizing, we should see that, for fixed $h$ (and hence fixed $L_s(h)$) and as we increase $L$, the derivative $\partial s_L/\partial L$ must change sign, going from positive to negative, and then approach zero from negative values. This is exactly what is observed in Fig.\ \ref{fig:dsdL}. The value of $L$ at which $\partial s_L/\partial L=0$ is proportional to $L_s$ (the constant of proportionality being very close to 1), as is the position of the minimum of $\partial s_L/\partial L$ (the constant here is close to $1.3$).
 
\section{Conclusions and perspectives. }
We analyzed the distribution of the entanglement spectrum of a single eigenstate of a spin-1/2 Heisenberg chain with random magnetic fields, identifying the finite-size corrections and the phase transition to an MBL phase. Our central result is the identification of three length scales: the first one $L_s(h)$ determines the minimum system size at which one can separate bulk and boundary effects; the other two $\ell_1$, $\ell_2$ are born at $L=L_s(h)$, but while the first one ($\ell_1$) diverges as $L\to\infty$ and defines the boundary effects, the second one ($\ell_2$) remains finite in that limit and determines the entanglement properties of local dynamics. We have developed new precision tools for the study of the distribution of the entanglement spectrum and we foresee that they will be useful for the analysis of the numerics of many other models, both disordered and clean. 

The study of the size dependence of the entanglement entropy of a finite region \cite{holzhey1994geometric,calabrese2004entanglement, Cardy:2010zs, doyon2009bipartite} can unveil crucial details about the critical theory underlying the MBL transition, which is, currently, unknown. {Although MBL transitions (in particular because of the infinite temperature) are not the natural setting where quantum field theory has been applied, an effective model which uses the emergent scale invariance at $h_c$ probably has a natural description in terms of QFT. If this is the case, the study of entanglement in MBL can be the back-door to develop a QFT of the MBL transition.}

\subsection*{Acknowledgements}
We would like to thank B.L.\ Altshuler, P.\ Calabrese, A.\ De Luca, D.A.\ Huse, F.\ Slanina and M.\ Schiulaz for useful discussions. FP would also like to thank ICTP for hospitality during the completion of this work. 
SP was partially supported by INFN through the project ``QUANTUM".
This project has received funding from the European Research Council (ERC) under the European Union’s Horizon 2020 research and innovation programme (grant agreement 694925).

\section*{References}
\bibliographystyle{iopart-num}

\bibliography{biblioMBL_proofread.bib}
 
\end{document}